\documentclass[12pt,a4paper]{article}
\usepackage{amsmath}
\usepackage{amssymb}
\usepackage[nosort]{cite}
\def\gfxon{\usepackage[final]{graphicx}}

\gfxon

\makeatletter
\@addtoreset{equation}{section}
\makeatother

\let\displaymath=\[
\let\enddisplaymath=\]
\def\[{\begin{equation}}
\def\]{\end{equation}}
\def\<{\begin{myeqnarray}}
\def\>{\end{myeqnarray}}

\makeatletter
\let\old@makecaption=\@makecaption
\def\@makecaption{\small\old@makecaption}
\makeatother

\makeatletter
\let\old@startsection=\@startsection
\renewcommand{\@startsection}[6]{\old@startsection{#1}{#2}{#3}{#4}{#5}{#6\mathversion{bold}}}
\makeatother

\newcommand{\hypref}[2]{\ifx\href\asklfhas #2\else\href{#1}{#2}\fi}

\newcommand{\figref}[1]{Fig.~\ref{#1}}
\newcommand{\Figref}[1]{Fig.~\ref{#1}}
\newcommand{\tabref}[1]{Tab.~\ref{#1}}
\newcommand{\secref}[1]{Sec.~\ref{#1}}
\newcommand{\indup}[1]{_{\mathrm{#1}}}

\newcommand{\sfrac}[2]{{\textstyle\frac{#1}{#2}}}
\newcommand{\half}{\sfrac{1}{2}}

\newcommand{\atopfrac}[2]{\genfrac{}{}{0pt}{}{#1}{#2}}

\newcommand{\Op}{\mathcal{O}}
\newcommand{\order}[1]{\mathcal{O}(#1)}

\newcommand{\superN}{\mathcal{N}}
\newcommand{\gym}{g_{\scriptscriptstyle\mathrm{YM}}}

\newcommand{\Tr}{\mathop{\mathrm{Tr}}}
\renewcommand{\Re}{\mathop{\mathrm{Re}}}

\newcommand{\cC}{{\cal C}}
\newcommand{\pint}{\makebox[0pt][l]{\hspace{3.4pt}$-$}\int}
\newcommand{\grSU}{\mathrm{SU}}
\newcommand{\grSO}{\mathrm{SO}}
\newcommand{\grO}{\mathrm{O}}

\newcommand{\lrbrk}[1]{\left(#1\right)}
\newcommand{\bigbrk}[1]{\bigl(#1\bigr)}

\newcommand{\nln}{\nonumber\\}

\newcommand{\nle}{\nonumber\\&=&\mathrel{}}
\newcommand{\eq}{\mathrel{}&=&\mathrel{}}
\newenvironment{myeqnarray}{\arraycolsep0pt\begin{eqnarray}}{\end{eqnarray}\ignorespacesafterend}
\newenvironment{myeqnarray*}{\arraycolsep0pt\begin{eqnarray*}}{\end{eqnarray*}\ignorespacesafterend}

\newcommand{\ads}{$AdS_5\times S^5$}

\newcommand{\eps}{\epsilon}

\newcommand{\OO}{{\cal O}}

\newcommand{\ben}{\begin{eqnarray}\displaystyle}
\newcommand{\een}{\end{eqnarray}}

\newcommand{\al}{\alpha}
\newcommand{\om}{\omega}

\newcommand{\s}{\sigma}
\newcommand{\ts}{\widetilde\sigma}
\newcommand{\la}{\lambda}

\begin{document}

\renewcommand{\thefootnote}{\fnsymbol{footnote}}

\thispagestyle{empty}
\begin{flushright}\footnotesize\tt
hep-th/0306139\\
AEI 2003-044\\
UUITP-09/03\\
ITEP-TH-31/03\\
\end{flushright}
\vspace{1cm}
\setcounter{footnote}{0}
\begin{center}
{\Large{\bf Stringing Spins and Spinning Strings \par}
    }\vspace{7mm}
{\sc N. Beisert$^a$, J.A.~Minahan$^b$,  
M. Staudacher$^a$ 
and K. Zarembo$^{b,}$\footnote{also at ITEP, Moscow, Russia.}
\\[7mm]
\it ${}^a$ Max-Planck-Institut f\"ur Gravitationsphysik\\
Albert-Einstein-Institut\\
Am M\"uhlenberg 1, D-14476 Golm, Germany}\\[2mm]
{\tt nbeisert,matthias@aei.mpg.de}\\[6mm]
{\it ${}^b$ Department of Theoretical Physics\\
Uppsala University\\
Box 803, SE-751 08, Uppsala, Sweden}\\[2mm]
{\tt joseph.minahan,konstantin.zarembo@teorfys.uu.se}\\[20mm]

{\sc Abstract}\\[2mm]
\end{center}

We apply recently developed integrable spin chain and dilatation 
operator techniques in order to compute the planar 
one-loop anomalous dimensions
for certain operators containing a large number of scalar fields
in $\superN =4$ Super Yang-Mills. The first set of operators, belonging to
the $\grSO(6)$ representations $[J,L-2J,J]$, interpolate smoothly
between the BMN case of two impurities ($J=2$) and the extreme case
where the number of impurities equals half the total number of
fields ($J=L/2$). The result for this particular $[J,0,J]$ operator
is smaller
than the anomalous dimension derived by Frolov and Tseytlin [hep-th/0304255]
for a semiclassical string configuration which is the dual of
a gauge invariant operator in the same representation.
We then identify a second set of operators
which also belong to $[J,L-2J,J]$ representations, but which
do not have a BMN limit.  In this case the anomalous dimension of
the $[J,0,J]$ operator does match the Frolov-Tseytlin
prediction.  We also show that the fluctuation spectra for this
$[J,0,J]$ operator is consistent with the string prediction.

\newpage
\setcounter{page}{1}
\renewcommand{\thefootnote}{\arabic{footnote}}
\setcounter{footnote}{0}

\section{Introduction}  
\label{sec:intro}

In the \ads{} conjecture \cite{Maldacena:1998re,Gubser:1998bc,Witten:1998qj}, 
it was argued that the anomalous dimension
of a generic gauge invariant operator scales
as 
\[
\label{gammastr}
\gamma\sim (\la)^{1/4}\sqrt{\ell},
\]
 where $\lambda=\gym^2N$ is the 't Hooft coupling and 
$\ell$ is the string level \cite{Gubser:1998bc}.
However, there are some operators which do not behave like \eqref{gammastr}
even for large values of $\lambda$ \cite{Polyakov:2001af}.
One example are the half-BPS operators whose
dimensions are protected by supersymmetry.  

Another example are the BMN operators,
which are nearby to the half-BPS operators~\cite{Berenstein:2002jq}.
These operators have been heavily studied 
\cite{Kristjansen:2002bb,Gross:2002su,Constable:2002hw,Santambrogio:2002sb,
Parnachev:2002kk,Gursoy:2002yy,Beisert:2002bb,Gross:2002mh,Constable:2002vq,
Eynard:2002df,Beisert:2002tn,Gursoy:2002fj,Minahan:2002ve,Beisert:2002ff,
Klose:2003tw,Beisert:2003tq,Freedman:2003bh}.
For these operators, one starts with a half-BPS operator 
$\Tr(Z^J)$
where $Z=\phi_5+i\phi_6$,
and then inserts ``impurities''.   
The impurities, e.g. the scalars $X=\phi_1+i\phi_2$ 
and $Y=\phi_3+i\phi_4$, are inserted in various positions in the
operator
\[\label{ops}
\OO=\Tr Z\ldots Z\,X\,Z\ldots Z\, X\, Z\ldots Z\, Y\, Z\ldots Z
\]
Alternatively, one can think of this as an operator with $L$ scalar fields, 
where $L-J$ of the original $Z$'s are converted to one of the other scalars. 
One-loop effects mix the operators among themselves and the
operators with definite scaling dimension, along with
their anomalous dimensions, are obtained by diagonalizing a matrix. 
This matrix of anomalous dimensions is 
particularly easy to compute by analyzing the 
action of the dilatation operator
\cite{Beisert:2002ff,Beisert:2003tq}, 
even in the non-planar sector and at higher-loops.
  
In the BMN limit, where the number of impurities is small
compared to $L$, it was shown that the eigenstates of this dilatation
operator are isomorphic to  particles on a circle of length $L$, 
with the number of particles
equal to the number of impurities.  With a small number of impurities
one can safely ignore the particle interactions
and use the dilute gas approximation.  The total momentum of the particles
has to be zero in order to guarantee cyclicity of the trace, and one finds
that the one-loop anomalous dimension behaves like
\[
\label{BMNad}
\gamma=\frac{\lambda}{8\pi^2}\sum_i\eps(p_i)=\frac{\lambda}{8\pi^2}\sum_i(p_i)^2
\]
where $p_i$ are the individual particle momenta and $\eps(p_i)$ can be thought
of as their energies.  Quantization on a circle of length $L$ then gives
\[
\label{BMN2}
\gamma=\frac{\lambda}{2L^2}\sum_i (n_i)^2
\]
where the $n_i$ are integers.

Remarkably, these results can be directly related to  a string calculation
on \ads{} \cite{Metsaev:2001bj,Metsaev:2002re,
Spradlin:2002ar,Verlinde:2002ig,Spradlin:2002rv,
Pankiewicz:2002gs,Zhou:2002mi,Vaman:2002ka,Pearson:2002zs,Gomis:2002wi,Pankiewicz:2002tg,
Roiban:2002xr,Gomis:2003kj,Spradlin:2003bw}, 
since one can choose $\lambda$ large but $L$ also large such
that $\la'=\la/L^2$ is small.  One then finds a one to one map of
the BMN operators to string states propagating in the plane wave limit 
\cite{Blau:2001ne,Blau:2002dy} of \ads{}.  
These states can be written as
\[
\prod_i (\alpha^{\mu_i}_{n_i})^\dagger|p_+\rangle
\]
where $p_+=L$ in appropriate units.  In the limit of small $\lambda'$, the
results then match with \eqref{BMN2}.

More recently, it has been suggested that other operators can
be compared with semiclassical string 
calculations \cite{Gubser:2002tv,Frolov:2002av,Russo:2002sr,Minahan:2002rc,Tseytlin:2002ny,Frolov:2003qc}.
The operators of \cite{Minahan:2002rc,Frolov:2003qc}
have order $L$ impurities, with each impurity carrying 
momentum of order $1/L^2$.  
The anomalous dimension is now of order $\lambda/L$.
The corresponding string states are expected to
have order $L$ creation operators.

In particular, Frolov and Tseytlin considered operators that
have $J'$ of the $Z$ scalars and an equal number $J$ of the 
$X$ and $Y$ scalars \cite{Frolov:2003qc}.
Assuming that these are highest weight states, one finds that these
are operators in the $[0,J'-J,2J]$ representation of $\grSO(6)$
if $J'\ge J$, or the $[J-J',0,J+J']$ representation if $J>J'$, up
to conjugation.
In the extreme case where $J'=0$, one finds a semiclassical result which
is twice the BMN prediction \cite{Frolov:2003qc}.  
That there is a difference from BMN is not surprising -- 
the density of  impurities is of order 1, so the dilute
gas approximation is naively expected to break down.  

The case of a very large operator with a high density of impurities
is not easily accessible by field theoretic means.
Nevertheless we can still make progress 
in computing the one-loop planar anomalous dimension for 
these types of operators. 
This is because the planar one-loop dilatation operator for scalars,
representing the interaction found in \cite{Beisert:2002bb},
is isomorphic to the Hamiltonian for an integrable system \cite{Minahan:2002ve}
and one can then use all of the machinery that goes along with it. 

In particular, if one restricts the operators to be composed only of
$Z$ and $X$ scalars, then the problem simplifies dramatically%
\footnote{Operators with only $Z$ and $X$ scalars are 
special as they only mix amongst themselves due to charge conservation
\cite{Beisert:2003tq}.}.
This is because the dilatation operator 
can be mapped directly to the Hamiltonian of the
Heisenberg spin chain, which was solved long ago by Bethe.  
The $Z$ scalars inside $\OO$ are the up spins and the $X$ scalars are 
the down spins.  
The field theory representation $[J,L-2J,J]$ corresponds to a 
configuration of total spin $L/2-J$.
The eigenstates for the spin chain can be found by solving the Bethe equations.
These are easy to solve for the case of two spin flips%
\footnote{In the context of $\superN=4$ SYM
these are the finite $J$ versions of the two-impurity BMN operators
which were identified and diagonalized in \cite{Beisert:2002tn}.},
but for more than two flips these are in general only solvable numerically.

Solving for a finite number of $L/2$ spin flips with total spin $0$ 
looks almost hopeless.
However, in the large $L$ limit, there is some hope of making progress
since here one can convert the Bethe equations to an
integral equation.  
This is how the ground state of the \emph{antiferromagnetic} Heisenberg
spin chain was solved; 
it has total spin $0$ and energy proportional to $-L$. 

To make contact with the proposal of \cite{Frolov:2003qc}
the low-lying states 
of the \emph{ferromagnetic} spin chain
with total spin $0$
and energy proportional to $1/L$
are of particular interest. 
Here, the antiferromagnetic ground state is the state 
with the highest energy. 
In contrast, 
the semiclassical string configuration is expected to correspond
to the state of the lowest energy, or at least much lower energy.

In this paper we solve this problem 
for $[J,L-2J,J]$ representations 
where $J$ and $L$ are assumed to be large. 
This corresponds to operators of the form (modulo ordering of fields)
\[\label{ops2}
\OO=\Tr Z^{L-J} X^{J} + \ldots
\]
As in the solution for the antiferromagnet ground state, this is solved
by converting the Bethe equations to an integral equation.  However,
unlike the case of the antiferromagnetic ground state, 
the Bethe roots do not lie
on the real line, but instead 
extend into the complex plane.  

We consider two classes of $[J,L-2J,J]$ 
operators.  The first class 
has two contours of Bethe roots symmetric
about the imaginary axis which we call the double contour solution.
The second class 
has all Bethe roots on the imaginary axis which we
call the imaginary root solution.  These latter Bethe states are 
``singular'' in the sense
that the Bethe equations need to be regularized.  
One of the main results of this paper
is to identify the $[J,0,J]$ (spin $0$) imaginary root
solution 
 as the SYM dual of the string solution found by Frolov and Tseytlin
\cite{Frolov:2003qc}.

The agreement between the SYM  and semiclassical string theory calculations
is remarkable: the anomalous dimension exactly
coincides with the energy of the string, and the spectrum of 
nearby operators coincides with the spectrum of small fluctuations
around the classical string solution.  Furthermore, the semiclassical
analysis indicates that there is an instability in the string solution.
This is consistent with our analysis, where we show that the imaginary
root solution is not the lowest energy $[J,0,J]$ state.  Instead we show 
that the 
double contour solution is the ground state for this
representation\footnote{Strictly speaking, the double contour solution is
valid only for even $J$ while the imaginary root solution is valid only
for odd $J$.  Indeed, the ground state for odd $J$ has
its Bethe roots in a completely different configuration than the double contour, 
and we have yet to determine the even $J$ analog of the imaginary root
solution.  However, in the thermodynamic limit (large $L$), there should
be no distinction between even and odd and we will provide strong
evidence for this in \secref{sec:comparison}.}.

We first consider the double contour Bethe state.  In the limit of small $J$,
this state approaches the BMN states with $J/2$ each of $\alpha_{-1}$ 
and $\alpha_{+1}$ oscillators.
We are able to solve this problem by first
considering the unphysical region where $J<0$, and then analytically
continuing to $J>0$. The problem then reduces to solving for two
elliptic equations which is easily done numerically
to any desired precision. For the
case where $J=L/2$, and in the strict $J \rightarrow \infty$ limit,
we find that the anomalous dimension is given by
\[
\label{upshot}
\gamma=\frac{\la J}{L^2}~\beta=\frac{\la}{4J}~\beta, \qquad\mbox{with }\beta=0.7120321458...
\]
The result in \eqref {upshot} is bigger than the BMN result 
\eqref{BMN2}
where one finds $\beta=1/2$.  However, \eqref{upshot}
is smaller than the 
semiclassical string result $\beta=1$ in \cite{Frolov:2003qc}.  Hence,
we conclude that this double contour solution is {\it not} the gauge
dual to the semiclassical string.

To help us find the gauge dual to the semiclassical string,
we compare our result in \eqref{upshot} to actual numerical computations
of $[J,0,J]$ ground states. 
We find very close agreement when $J$ is even.  However,
when $J$ is odd, we find that while the result in \eqref{upshot} appears
correct, the configuration of Bethe roots is completely different from the
even case.  In this case there are two curving contours in the complex
plane connected by a straight line of roots along the imaginary axis.
It would be interesting to solve the integral equation describing this
distribution of Bethe roots.

Nevertheless, 
it turns out that there are also solutions where all
Bethe roots lie on the imaginary axis.  
With all imaginary roots, the relative coefficients in the sum over states for the spin 
chain (or gauge invariant operator) are real.
For $J$ small compared to $L$, the roots form two overlapping ``strings''
where the separation between roots is very close to $i/2$.  If $J$ is
of the same order as $L$, then the outer roots spread out.
If we then take the limit
of large $J$ and $L$, we can reduce the Bethe equations to an integral
equation which is very similar to the equation found by Douglas and
Kazakov for two-dimensional QCD on the sphere 
\cite{Douglas:1993ii,Kazakov:1996ae}.
We find that when $J=L/2$ there is a critical point and for this value
the anomalous dimension
matches the prediction in \cite{Frolov:2003qc}.  Unlike the case of
the double contour, as the number of roots is reduced, the state
{\it does not} approach a BMN state.  

To further verify that this imaginary root solution is the gauge dual of the
semiclassical string,
we consider spinless fluctuations about the solution.  This
is done by pulling roots off of the imaginary axis and onto (or close to)
the real axis.  Here we find a spectrum that is consistent with that
found in \cite{Frolov:2003qc}.

In section two we solve for the solution with a double contour of roots
in the large $L$ limit.  We find $\gamma$ as a function of $L$ and
show that it gives \eqref{upshot}. 
In section three we compare this result to actual numerical computations
of $[J,0,J]$ states.  Here we show the surprising difference between
the even and odd configurations. In section four we consider and solve for the
imaginary root solution in the large $L$ limit.  We show that it matches
the predictions in \cite{Frolov:2003qc} for the anomalous dimension
and the fluctuation spectrum.  In section five we consider
an analogous solution to an $\grSO(6)$ singlet configuration and show that
it matches a semiclassical prediction made in \cite{Minahan:2002rc}.
In section six we present our conclusions.
 
\section{The double contour solution}
\label{sec:2contour}

In this section we describe the ground state for the $[J,L-2J,J]$ representations.  Strictly speaking, this solution is valid only if $J$ is even.  If $J$ is
odd and $L>2J$, 
then the true ground state
is doubly degenerate.  But it approaches the double contour
solution in the limit of large $L$.  In the next section, we will argue
that if $J$ is odd and $L=2J$, then the ground state will have a 
completely different root configuration than the solution described in
this section.  However, the anomalous dimension appears to be the same.

We begin with the Bethe equations, which read\footnote{For a nice review
see \cite{Faddeev:1996iy}.} 
\[
\label{ansatz}
\left( \frac{u_j+i/2}{u_j-i/2} \right)^L=
\prod_{\atopfrac{k=1}{k\neq j}}^{J} \frac{u_j-u_k+i}{u_j-u_k-i}
\]
Here $L$ is the total length of the spin chain and $J$ is the
number of impurities\footnote{Note that this
connotation has recently been introduced in the BMN literature.
This is different from impurities in condensed matter theory,
where one would rather speak, equivalently, of ``excitations'',
``particles'' or ``magnons''. Or even simpler, ``spins down''!}.
The complex numbers $u_j$ are the rapidities of the upside down spins 
($2u_j=\cot p_j$, where $p_j$ is the momentum of $j$th impurity). 
 Normally the rapidities are real, but the impurities
can form bound states in which case their rapidities acquire imaginary 
parts. In the literature on the Bethe ansatz the bound states of impurities
are usually referred to as ``strings''. We shall study bound states of 
a macroscopically large number of impurities and shall argue that
these large ``strings''
in the Heisenberg model 
are dual to strings in $AdS_5\times S^5$
spinning around the five-sphere.
The state with $J$ impurities belongs 
to the representation $[J,2L-J,J]$ of $\grSO(6)$ and we
define the filling fraction 
\[
\label{filling}
\alpha=\frac{J}{L}\le\frac{1}{2}.
\]
We can also write down the logarithmic version of the Bethe equations
\[\label{logansatz}
L\, \log \frac{u_j+i/2}{u_j-i/2} =
2\pi i n_j+ \sum_{\atopfrac{k=1}{k\neq j}}^{J} \log\frac{u_j-u_k+i}{u_j-u_k-i}
\]
where the numbers $n_j$ reflect the ambiguity of the branches of
the logarithm.
These numbers can be interpreted as the lattice momentum
of the $j$-th `particle'.

Since the spin chain corresponds to an operator with a
trace, we are only interested in states that are invariant under
shifts.  This forces us to set the trace condition
\[\label{trcond}
\prod_{j=1}^{J}\frac{u_j+i/2}{u_j-i/2}\ =\ 1. 
\]
The eigenvalues of the planar dilatation operator can be found by
solving the system of Bethe equations for the roots
$u_j$ ($j=1, \ldots, J$) \cite{Minahan:2002ve}.
This is somewhat cumbersome for chains of small length,
but very advantageous for long chains where a direct diagonalization
\cite{Beisert:2002ff,Beisert:2003tq} is no longer feasible.
The planar anomalous dimensions are then given to one-loop, 
for general $L$ and $J$, by
\[
\label{anomal}
\Delta=L + \gamma \qquad {\rm with} \qquad
\gamma=\frac{\lambda}{8 \pi^2} \sum_{j=1}^{J}
\frac{1}{u_j^2+1/4}\ 
\]
In the ``thermodynamic limit'' where $L \rightarrow \infty$, the roots are of
order $u_j=\order{L}$. Let us therefore rescale them
as
\[
u_j \rightarrow L~u_j
\]
Then one easily proves that the Bethe equations 
\eqref{logansatz} 
simplify
for $L \rightarrow \infty$ to
\[
\label{sumequation}
\frac{1}{u_j}=2 \pi n_j+ 
\frac{2}{L} \sum_{\atopfrac{k=1}{k \neq j}}^{J}
\frac{1}{u_j - u_k} .
\]
Here, the term on the left corresponds to a global potential,
while the second term on the right 
corresponds to pairwise repulsion of roots.
As is standard when taking the thermodynamic limit 
of the Bethe equations, 
eq.~\eqref{sumequation} can be  conveniently turned
into an integral equation. 

In order to turn the Bethe equations \eqref{sumequation}
into a single integral equation we need to understand
the structure of the distribution of Bethe roots in the
complex plane. To gain some intuition, let us
concentrate on the case that $L$ and $J$ are large but the
filling fraction $\al\ll 1/2$. 
We will also assume for now that $J$ is even. 
The trickier case of odd $J$ will be discussed (but not analyzed to
the end) below.
As can be seen from \eqref{sumequation}, to a reasonable approximation the 
Bethe roots may be placed  around the positions $u_i=1/(2\pi n_i)$, as 
long as no two $n_i$ are the same. 
However, we wish to lower the energy as much as possible, which means
setting  all $n_i$ to $\pm1$ with an equal number of each to 
satisfy the condition in \eqref{trcond}.  
The interaction term in \eqref{sumequation}
pushes the roots apart.  It is well known that the roots
are pushed in the imaginary direction and form ``strings"
roughly parallel to the imaginary axis
with the separation between the
$u_i$ of order $1/\sqrt{L}$ \cite{Minahan:2002ve}.  It is clear from the Bethe equations
that given a root $u_i$, there has to be a root $u_j=u_i^*$, therefore
the distribution of roots is invariant under reflection about the
real axis.  

\begin{figure}\centering
\includegraphics{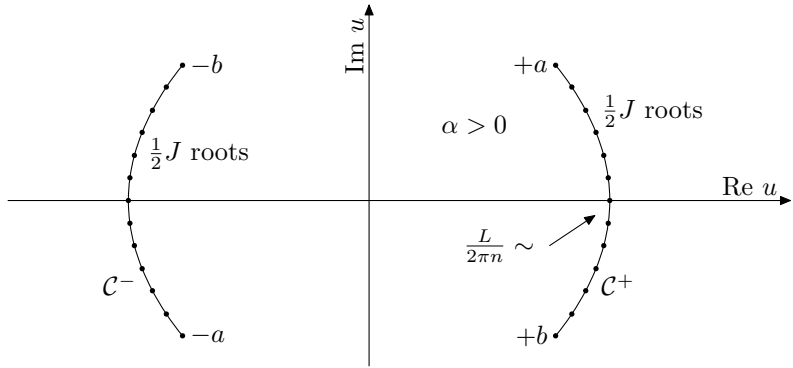}
\caption{Bethe roots. For large $L$ the roots condense into two cuts}
\label{fig:roots}
\end{figure}

It was also shown that there was  a $1/L$  correction
in the real direction towards the imaginary axis, 
where the correction is
larger the further the root is from the real line.  Finally, it was argued
that placing the roots like this led to corrections to $\gamma$
which were of order $1/L^3$, while changing the branch gives a $1/L^2$
correction.  Hence, at least for small $J/L$, it is energetically favorable
to evenly distribute the roots on the $\pm 1$ branches and hence the
distribution looks
like that in \Figref{fig:roots}.   

If $J$ is of order $L$, then  adding a root to the heavily populated
$\pm1$ branches costs an energy of order $1/L^2$.  However,  we will assume
that for $\alpha\sim 1/2$ it is still favorable to place the roots
in these branches.  We can also make a slight generalization and
place all of the roots in the branches $\pm n$.  From the
semiclassical string perspective, we expect this to correspond to a
string that winds $n$ times around itself.  The net effect of this is
to multiply $\sqrt{\lambda}$, the effective string tension, by a factor
of $n$.

\begin{figure}\centering
\includegraphics{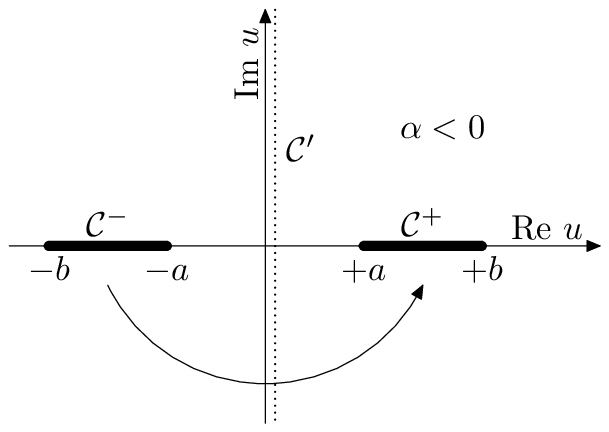}
\qquad
\includegraphics{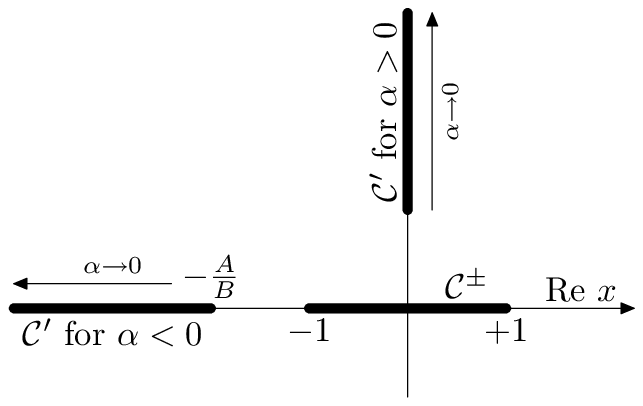}
\caption{Cuts flip for $\alpha<0$.
Both cuts can be mapped to one $\mathcal{C}^\pm$ via the map
$u^2=A+Bx$.}
\label{fig:cuts}
\end{figure}

Assuming that $J$ is still even, the roots are distributed so that
they are symmetric under reflection about
the imaginary axis.  Hence we only need to consider the
first $J/2$
roots $u_j$ with $j=1, \ldots, J/2$, since we can now assign 
$u_{j+J/2}=-u_j$.  Based on our previous argument, we
now assume that $n_j=n=1, 2,\ldots$. Let us introduce a
density
\[
\label{density}
\rho(u) = \frac{2}{J} \sum_{k=1}^{J/2} \delta(u-u_k)
\]
describing the distribution of roots with positive real part
in the complex plane.
This is expected to turn into a smooth function in the
limit of large $L$, with a continuous support tracing a 
curve $\cC^+$ (see \Figref{fig:roots}).
The density is normalized to one:
\[
\label{norm1} 
\int_{\cC^+} du~\rho(u)=1
\]
One easily verifies that the Bethe equations \eqref{sumequation}
can be reexpressed with the help of the density as
\[
\label{inteq1}
\pint_{\cC^+} dv~\rho(v)~\frac{v^2}{u^2-v^2} =
\left(\frac{1}{2 \alpha}-1 \right) -\frac{\pi n}{\alpha}~u 
\]
Here the integral has to be understood in the principal part
sense, as we need to omit the case $k=j$ in equation
\eqref{sumequation}. The anomalous dimension, {\it cf}
eq.~\eqref{anomal} becomes (remember the rescaling 
$u_j \rightarrow L u_j$)
\[
\label{anomal2}
\gamma=\frac{\lambda}{8 \pi^2} \frac{\alpha}{L} 
\int_{\cC^+} du~\frac{\rho(u)}{u^2}
\]
Therefore all we need to do is invert the singular integral
equation \eqref{inteq1}; this yields the density, and therefore,
in light of eq.~\eqref{anomal2}, the anomalous dimension.
Luckily, similar equations have appeared previously in the
case of large $N$ matrix models (which, incidentally, are also
integrable models of sorts). In particular, equation 
\eqref{inteq1} is a close relative of the equations describing
the solution of Kostov's $\grO(n)$ multi-matrix model\footnote{
see \cite{Kostov:1992pn} and references therein.}.
An inconvenient feature of \eqref{inteq1} is that
the precise shape of the contour ${\cal C}^+$ is unknown.
The situation can be improved by
first solving the equation in the unphysical regime $\alpha<0$,
corresponding to a negative number of down spins.
Here we expect the cuts to lie on the real axis, as is shown in the
matrix model problems. The contour $\cC^+$
becomes the real interval $[a,b]$ while the mirror contour $\cC^-$
turns into the flipped interval $[-b,-a]$, as in the left half of
\Figref{fig:cuts}.
This way we merely need to determine the endpoints $a,b$ but
not the shape of the contour. After solving the equation
for general negative $\alpha$ we can subsequently analytically
continue back to the physical regime $\alpha \in [0,1/2]$.
Next, let us apply the same folding map as in the case of the
$\grO(\pm 2)$ matrix model \cite{Kostov:1992pn}:
\[
\label{map}
u=\sqrt{A+B x} \quad {\rm with} \quad A=\half (b^2+a^2)
\quad {\rm and} \quad B=\half (b^2-a^2)
\]
This maps the two cuts $[a,b]$ and $[-b,-a]$ on top of
each other, and to the interval $[-1,1]$.
Our equation becomes, 
\[
\label{inteq2}
\pint_{-1}^{1} dy \,\rho(y) \frac{\sqrt{A+B y}}{x-y} =
\left( \frac{1}{\alpha}-2 \right)- \frac{2 \pi n}{\alpha} \sqrt{A+B x}
\]
where $\rho(u)\equiv\rho(x(u))$.  
Equation \eqref{inteq2} is a standard one-matrix model 
(or airfoil) equation
with a non-polynomial potential involving $\sqrt{A+B x}$. The new
cut $\cC'$ with branchpoints at $x=\infty$ and $x=-A/B$ appears
due to the above folding map, as shown in the right part of
\Figref{fig:cuts}. The latter 
branchpoint lies for $\alpha<0$ to the left of $-1$
on the negative real axis, while for $\alpha>0$ it sits
on the imaginary axis. The solution for the density is
immediately obtained by an inverse finite Hilbert transform:
\[
\label{solution1}
\rho(x)=
-\frac{1}{\pi^2} \frac{\sqrt{1-x^2}}{\sqrt{A+B x}}
\pint_{-1}^1 dy \frac{1}{x-y} \frac{1}{\sqrt{1-y^2}}
\left[ 
\left( \frac{1}{\alpha}-2 \right)- \frac{2 \pi n}{\alpha} \sqrt{A+B y}
\right].
\]
It simplifies to
\[
\label{solution2}
\rho(x)=
\frac{2 n}{\pi \alpha}\, \frac{\sqrt{1-x^2}}{\sqrt{A+B x}}
\pint_{-1}^1 dy\, \frac{1}{x-y}\, \frac{\sqrt{A+B y}}{\sqrt{1-y^2}}.
\]

This is nearly the solution to our problem, except that
we still need to determine
the parameters $A,B$ as a function of the filling fraction
$\alpha$. These are determined from the normalization condition
\eqref{norm1} which reads in the new variables
\[
\label{norm2}
\frac{B}{2} \int_{-1}^1 dx\, \frac{\rho(x)}{\sqrt{A+B x}}=1
\]
and the further condition 
\[
\label{positivity1}
\int_{-1}^1 dx\, \frac{1}{\sqrt{1-x^2}}
\left[ 
\left( \frac{1}{\alpha}-2 \right)- \frac{2 \pi n}{\alpha} \sqrt{A+B x}
\right]=0
\]
which ensures the positivity of the density on the cut.
It simplifies to
\[
\label{positivity2}
\int_{-1}^1 dx\, \frac{\sqrt{A+B x}}{\sqrt{1-x^2}}=
\frac{1}{n} \left( \frac{1}{2}-\alpha \right).
\]
This completes the solution: $A$ and $B$ are determined 
through the two equations \eqref{norm2}, \eqref{positivity2},
which one could express through elliptic integrals of the
first and second kind, while the density is given by
equation \eqref{solution2}, which could be found explicitly in terms
of an elliptic integral of the third kind.
However, as we are chiefly interested in the anomalous dimension
\eqref{anomal2}, we can simply eliminate the density.
Plugging \eqref{solution2} into both the normalization condition
\eqref{norm2} and the expression for the anomalous dimensions
\eqref{anomal2}, we respectively find, 
after exchanging orders of integration,
\[
\label{norm3}
\int_{-1}^1 dx\, \frac{1}{\sqrt{1-x^2} \sqrt{A+B x}}=
\frac{1}{2 n} \frac{1}{\sqrt{A^2-B^2}}
\]
and
\[
\label{anomal3}
\gamma=\frac{\lambda}{16 \pi^2 L}
\left(
\frac{1-2 \alpha}{\sqrt{A^2-B^2}} - \frac{A}{A^2-B^2} 
\right)\ .
\]
Thus $\gamma$ is explicitly given by the parameters $A,B$ which
in turn are implicit functions of $\alpha$ and $n$,
as stated in eqs.~\eqref{positivity2}, \eqref{norm3}.
It is now a straightforward exercise to invert these equations
in terms of a series in $\alpha$ to any desired order, one finds
\[
\label{expansion}
\gamma=\frac{\lambda n^2 J}{2 L^2} 
\left
(1+\sfrac{1}{2}\alpha + \sfrac{3}{8} \alpha^2+\sfrac{21}{64} \alpha^3+
\sfrac{159}{512} \alpha^4 + \sfrac{315}{1024} \alpha^5
+\sfrac{321}{1024} \alpha^6 \ldots
\right)
\]
While we derived this result for negative $\alpha$, we see that
the function is perfectly analytic at $\alpha=0$ and we can
immediately analytically continue to the correct Bethe phase
$\alpha \in [0,1/2]$. 
The leading order $\order{\alpha}$ result correctly reproduces 
the dilute gas approximation \eqref{BMN2}
and the famous BMN result $\lambda n^2/L^2$ for $J=2$.
Let us also note that the leading $1/L$
correction to the BMN limit
\[
\gamma=\frac{\lambda n^2 J}{2L^2}\left(1+\frac{J+2}{2L}+\order{1/L^{2}}\right)
\]
was obtained in \cite{Minahan:2002ve};
it agrees with the $\order{\alpha^2}$ term in \eqref{expansion}
and thus yields a nice check of our method. The above series
\eqref{expansion} converges rather rapidly in the interval
$\alpha \in [0,1/2]$ allowing us to obtain the anomalous dimension
to arbitrary precision for any $\alpha$. 
We can also expand the anomalous dimension around $\alpha=1/2$, where
we find
\begin{equation}
\gamma=\,\frac{\lambda}{2L}\left[0.7120-1.0916(1-2\alpha)+\ldots\right]
\end{equation}
 A plot of these
results in presented in 
\Figref{fig:plot} in the next section. 

An alternative method for solving \eqref{positivity2} and \eqref{norm3} is
to analytically continue $\al$ first, thus moving $B$ onto the imaginary
axis.  For small values of $\alpha$, $A$ is positive, but decreases as 
$\alpha$ is increased and
eventually changes sign for a critical value of $\alpha$.  At this value
the branch point on the imaginary axis touches the cut along the real line.

To avoid this collision, we simply deform the integration on the real
line to an integration along the cut on the imaginary axis.  If we
define $C=-iB$, then the equation in \eqref{norm3} becomes
\[\label{norm4}
2\int_{A/C}^\infty dy\,\frac{1}{\sqrt{1+y^2}\sqrt{Cy-A}}=\frac{1}{2 n} \frac{1}{\sqrt{A^2+C^2}}.
\]
Defining a new variable $\om=\sqrt{Cy-A}$, \eqref{norm4} can be reexpressed
as
\[\label{norm5}
\int_{-\infty}^{\infty}d\om\,\frac{1}{\sqrt{(\om^2+A)^2+C^2}}=\frac{1}{2 n} \frac{1}{\sqrt{A^2+C^2}}\ .
\]
We then do the same with \eqref{positivity2},  although now there will
be a divergent piece that needs to be subtracted off.  Deforming the contour,
redefining variables and subtracting off the infinite piece, we
find 
\[\label{positivity3}
\int_{-\infty}^{\infty}d\om\left(1-\frac{\om^2}{\sqrt{(\om^2+A)^2+C^2}}\right)=\frac{1}{n} \left( \frac{1}{2}-\alpha \right).
\]

If we set $A$ to zero, then the elliptic integrals reduce to ordinary
integrals and we can find an analytic solution.  In this
case
\<\label{alcrit}
\alpha\eq\frac{1}{2}-\frac{4\pi^2}{(\Gamma(1/4))^4}\approx 0.2715 < \frac{1}{2},
\nln
\gamma\eq\frac{\la n^2}{2\pi L}.
\>
Hence $A$ becomes negative before $\al$ reaches $\al=1/2$.
Assuming that $A<0$ and letting $\eta=-C/A$ we define the integrals
\<
I_1(\eta)\eq\int_{-\infty}^{\infty}dz\left(1-\frac{z^2}{\sqrt{(z^2-1)^2+\eta^2}}\right)
\nln
I_2(\eta)\eq\int_{-\infty}^{\infty}\frac{dz}{\sqrt{(z^2-1)^2+\eta^2}}.
\>

We then find from \eqref{norm5} and \eqref{positivity3} that
\[
\al=\frac{1}{2}\left(1-\frac{1}{\sqrt{1+\eta^2}}\,\frac{I_1(\eta)}{I_2(\eta)}\right).
\]
While it is possible to solve this for all physical values of $\al$,
let us only consider the case
$\al=1/2$, which corresponds to setting $I_1$ to
zero.  This is easily done numerically, where one finds
$\eta=1.16220056$.  From this it is straightforward to show that the
value for $\gamma$ is that in \eqref{upshot} and that the
values for $A$ and $C$ are
\[
A=-0.086987288\qquad\qquad C=0.10109667.
\]
In terms of our original variable $u$, we see then that the branch points
are at
\[
\label{branchpts}
u=\pm 0.1523\pm 0.3319\ i.
\]

The exact distribution of the roots is found by starting
from the branch points  and finding the contour where $du\,\rho(u)$ is 
positive definite. 
The positivity of the density
 unambiguously determines the shape of the contour \cite{David:1991sk}.
Here, $\rho(u)$ is the analytic continuation 
of \eqref{solution2}.
In terms 
of $x$, this means 
\[
\label{rootdis}
\frac{B\rho(x)}{\sqrt{A+Bx}}\ dx
\]
is positive definite. While we have not solved for this over the whole
contour, we have determined that the slope of the contour away from
the upper right branch point in \eqref{branchpts} is $-0.743$.  Hence
the distribution has the general form in \Figref{fig:roots}.

\section{Comparison of results}
\label{sec:comparison}

We now compare our results of the previous section to 
the exact anomalous dimensions of 
operators of the form \eqref{ops2} for reasonably large values of $L$.
These can be obtained directly using the dilatation operator
or using the exact Bethe equations.

\subsection{Dilatation operator}
\label{sec:dilop}

For given $L$ and $J$ we collect all operators of the form
$\Tr Z^{L-J}X^J+\ldots$
and act on them with the dilatation operator \cite{Beisert:2003tq}.
In this way the matrix of anomalous dimensions is generated
and we subsequently diagonalize it.
A slight complication is that the set of operators \eqref{ops2}
comprises not only the operators in the representation
${[J,L-2J,J]}$, but also the ones for smaller values 
of $J$.
These additional eigenvalues have to be removed as
some of them are generically smaller than the one
we are interested in.
An easy way to achieve this is to
repeat the analysis for the operators 
$\Tr Z^{L-J+1} X^{J-1}+\ldots$
and remove the corresponding eigenvalues\footnote{
One of the benefits of the Bethe ansatz 
is that states of a lower representation are easily identified,
they have roots at infinity, and discarded.}.
What remains are exactly the 
anomalous dimensions of the
operators in the representation $[J,L-2J,J]$
\footnote{For the representation $[9,1,9]$ of $\grSU(4)$ this 
involved 4862 states. Out of these,
884 belong to the representation $[9,1,9]$, the other belong
to $[J,19-2J,J]$ with $J<9$.}.

\begin{figure}\centering
\includegraphics{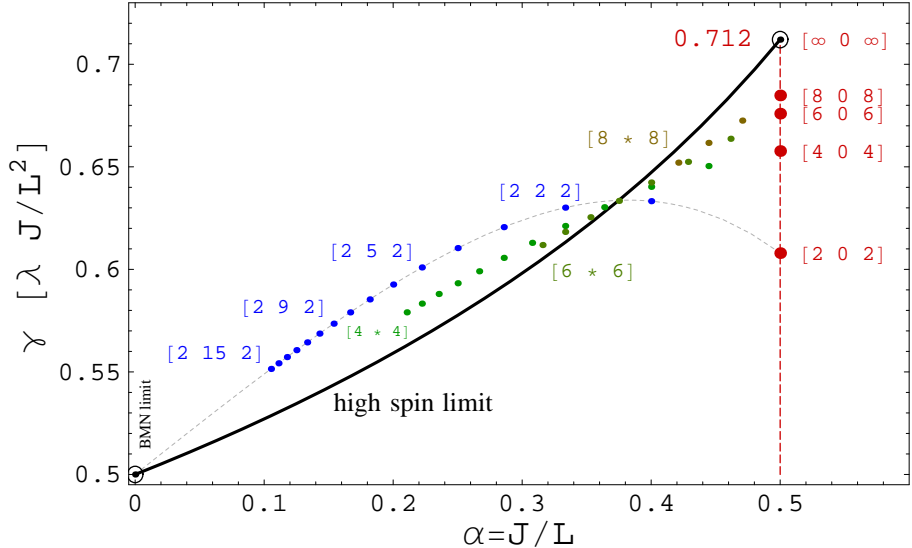}
\caption{
States of lowest energy in $[J,L-2J,J]$, $J$ even. 
The plot shows how the energy per spin flip increases from 
the dilute gas, $\alpha=0$, to the maximum filling $\alpha=1/2$. 
Discrete values for $L\leq 19$ are obtained as eigenvalues of the 
dilatation generator and the curve for $L\to\infty$ 
represents the solution to the Bethe equations.
The states with $J=2$ were found in \cite{Beisert:2002tn}.}
\label{fig:plot}
\end{figure}

In this way we have obtained the anomalous dimensions
for all operators of length $L\leq 19$.
We give the numerical values 
for the lowest eigenvalue operators in \tabref{tab:halffilled}
and \figref{fig:plot}.
We note that the behavior of the anomalous dimensions
is significantly different for even and odd values of $J$.
We will therefore treat them separately in what follows.

Already for $J=8$ the accuracy is very good
compared to the numerically exact result \eqref{upshot}
at $J\to\infty$. 
We can improve the estimate for $J\to\infty$ by
linearly expanding around $J=\infty$,
\[
\label{fitfunc}
\gamma=\bigbrk{a + b/ J+\ldots}\frac{\lambda J}{L^2}
\]
We fit the coefficients $a$, $b$ 
to the available data for $J=4,6,8$ and get
\[
\label{evenfit}
\gamma=\bigbrk{0.712 - 0.22/ J+\ldots}\frac{\lambda J}{L^2}\]
The estimates depend on the selected set of data and the 
fit function, overall the agreement 
with \eqref{upshot} seems to be within $1\%$.

\begin{table}\centering
$\begin{array}{|cc|cc|}\hline
J & \gamma\, [\lambda J/L^2] & J & \gamma\, 
[\lambda J/L^2]\\\hline
2&0.6079271019 & 3 & 0.9118906528\\{}
4&0.6576949106 & 5 & 0.8416441952\\{}
6&0.6759273535 & 7 & 0.8155060662\\{}
8&0.6848144787 & 9 & 0.7973649389\\\hline
\end{array}$
\caption{Numerical values for the lowest eigenvalues
in $[J,0,J]$}
\label{tab:halffilled}
\end{table}

Unlike the even case, we see from \tabref{tab:halffilled} that
for odd values of $J$ the energies seem to approach an asymptotic value of
$\order{\gamma L^2/J}$ from above.
A fit for odd $J$  with \eqref{fitfunc} yields
$\gamma=\lrbrk{0.743 + 0.49/ J}\lambda J/L^2$.
While the even and odd case seems to asymptote to a different value,
this could be due to the paucity of data points in the fit.
In the thermodynamic limit where $J$ is large, it is hard to 
believe there should be a distinction between even and odd.
Further down in this section, we will present evidence that indeed the
difference vanishes,
see \eqref{betteroddfit}.

We can make further observations on the obtained data for $L\leq 19$. 
An important feature of the one-loop planar dilatation generator 
or of integrable spin chains in general is that states
may come in pairs, see \cite{Beisert:2003tq}. 
We observe that 
\begin{list}{$\bullet$}{\itemsep0pt}
\item the energy of the ground state is $\order{1/L}$,
\item for even $J$ the ground state is unpaired,
\item for odd $J$ the ground state is paired unless $J=L/2$,
\item for odd $J$ the energy of the lowest unpaired state is $\order{1/J}$,
\item for odd $J$ and $L$ all states are paired.
\end{list}

It is also interesting to investigate the operators of lowest energy.
For $\alpha=1/2$ the operators are
\[
\Op=a \Tr Z^J X^J+ \sum_{m,n=1}^{J-1} b_{m,n} \Tr Z^m X^n Z^{J-m} X^{J-n}+\ldots
\]
where $a$ and $b$ are  mixing coefficients.
We find that $a$ is by far the most dominant coefficient.
The coefficients $b_{m,n}$ are less important, while the 
remaining coefficients are almost negligible.
This is in analogy to the Weiss domains configuration 
of a ferromagnet.
We find that alike spins cluster in domains
with only the domain walls contributing to the energy. 
For a minimal energy solution the number of domains must therefore 
be as small as possible.

\subsection{Bethe roots}

The Bethe equations are non-linear equations of
multiple variables $u_j$. It is therefore a difficult 
task to find solutions to them. 
A numerical solution requires the knowledge of 
the approximate position
of the roots $u_j$. It would therefore be useful 
to know how the roots arrange in general. 
We have worked out solutions for small values
of $J$ and $L$ to obtain some intuition in this
respect. 

First of all we note that the Bethe equations \eqref{ansatz},
the trace condition \eqref{trcond} and the 
energy \eqref{anomal} are invariant under 
negation of roots $u_j\to -u_j$.
Therefore each solution $\{u_j\}$ has a partner $\{-u_j\}$
of equal energy unless $\{u_j\}=\{-u_j\}$.
This is the same pairing discussed in the previous subsection.
Here we will concentrate on the unpaired states.
This is also simplifies the analysis because now there are only 
half as many equations to  solve.

For unpaired solutions we must distinguish between
an even and an odd number of roots $J$.
For even $J$ all roots come in pairs 
$u_{J/2+j}=-u_{j}$ and the trace condition 
\eqref{trcond} is automatically satisfied.
The important feature for odd $J$ is that
one of the roots must be zero,
while the others come in pairs.
At first glance, this would seem to violate 
the trace condition since the zero Bethe root contributes 
a factor $-1$ while $u_j$ and $-u_j$ together contribute a factor of $1$.

There is one way around this problem, which is to also place two Bethe roots
at the singular positions $\pm i/2$. These roots need to be regularized
since the Bethe equations diverge for these values%
\footnote{The existence of these singular solutions to the 
Bethe equations was
recognized by Bethe himself \cite{Bethe:1931hc} 
and has drawn certain attention with 
regard to the completeness of the Bethe ansatz 
\cite{Essler:1992wd,Kirillov:1997cp,Juttner:1994sk,Baxter:2001sx}.
}.
Hence we may assume that there are three roots 
\[
\label{singbet}
u_J=0,\qquad\qquad u_{J-1,J-2}= \ i(\pm 1/2+\varepsilon\pm \delta), 
\]
where $\delta=\order{\varepsilon^L}$. 
The Bethe equations associated to the singular roots 
and the trace condition are satisfied in the limit
$\varepsilon\rightarrow 0$. The 
Bethe equation associated to the zero root implies
that $L$ must be even. This is in agreement with 
the observation of the last subsection.
The contribution of the special roots to $\gamma$ is finite and is given by
\[
\gamma=\frac{\lambda}{8\pi^2}\lim_{\varepsilon\rightarrow 0}
\left(4-\frac{1}{\varepsilon+\varepsilon^2}-\frac{1}{-\varepsilon+\varepsilon^2}\right)
=\frac{3\lambda}{4\pi^2}\,.
\]
If $J=3$ then there are no other roots and this state 
corresponds to the special three impurity operators discussed in 
\cite{Beisert:2003tq}.

We have determined the root configurations for all
unpaired states of the $[J,0,J]$ spin chain for $2\leq J\leq 7$.
The positions of the roots were found by rewriting the Bethe 
equations in polynomial form.
These equations were combined into a single polynomial equation
of one variable using the resultant. All Bethe roots are then
among the solutions to this equation.
We summarize our findings in \tabref{tab:yetanothertable}
and \tabref{tab:yetanothertable606},~\ref{tab:yetanothertable707} at
the end of this paper.

\begin{table}\centering
$\begin{array}{|c|c|lll|c|}\hline
J&\gamma\, [\lambda J/L^2] & \multicolumn{3}{c|}{\mbox{Bethe roots}}&n \\\hline
2&0.607927& 0.288675             &         &   & 1 \\
\hline
3&0.911891& 0&\pm i/2&& 2\,\mathord{\ast}\\
\hline
4&0.657695&\multicolumn{2}{c}{(\pm 0.463265 \pm  0.502294i)}&& 1\,1\\
&1.104851&\pm 1.025705i&\pm 0.041309&& 0\,3\\
&2.290302&\pm 0.525012 &\pm 0.129473&& 1\,3\\
\hline
5&0.841644& 0&\pm i/2&\pm 0.998506i & 3\,\mathord{\ast}\,1\\
&1.291310& 0&\pm i/2&\pm 1.570673i & 4\,\mathord{\ast}\,0\\
&2.289413& 0&\pm i/2&\pm 0.638965  & 4\,\mathord{\ast}\,1\\
&3.176721& 0&\pm i/2&\pm 0.236124  & 4\,\mathord{\ast}\,2\\ 
\hline
\end{array}$
\caption{States of $[J,0,J]$}
\label{tab:yetanothertable}
\end{table}

We make the following observations. 
All the roots of a given mode number $n$ form a vertical string of roots
approximately at the real coordinate $L/2\pi n$.
This is in agreement with \cite{Minahan:2002ve}.
What is different is the distance of roots in the imaginary direction.
For a low density of roots $\alpha$, the roots of equal mode number
$n$ should be separated by $i\order{\sqrt{L}}$. For a high 
density the separation is very close to $i$ if the roots are close
to the real axis, but spreads out as we move away from the axis. 

\subsection{The odd ground state}\label{sec:oddground}

Since the three roots in \eqref{singbet} contribute a finite amount to
$\gamma$, the other roots must somehow cancel off most of this contribution
to leave a $\gamma$ of order $\order{1/L}$.  We note that if a Bethe root
is on the imaginary axis with $|u_i|>1/2$, then this root  contributes
a negative amount to $\gamma$.  

Let us examine more closely the Bethe roots for relatively small odd 
$J$.  Assuming the presence of the singular roots in \eqref{singbet}, the
$[5,0,5]$ state has two more roots which are the negative of each other
and which have to be either real or imaginary in order to be invariant under
the shift operator.  Hence, the Bethe equations in \eqref{ansatz}
reduce to a single equation
\[
\left(\frac{u+i/2}{u-i/2}\right)^8=\frac{(u+i)(u+3i/2)}{(u-i)(u-3i/2)}.
\]
One then finds that the solution that gives the entry in 
\tabref{tab:halffilled} is imaginary and very close to $i$.
A similar analysis for the ground state of the $[7,0,7]$ operator
 shows that two more roots are
placed very close to $\pm 3i/2$ and that the other two roots move even closer
to $\pm i$.  This pattern continues up to and including the $[11,0,11]$
operator, where roots are added near half integer imaginary values and where
the lower roots move closer to the half integers as higher roots are
added.

The half-integer periodicity of imaginary roots
can be explained by the same argument that leads to the string hypothesis
for the structure of Bethe states in the thermodynamic limit 
\cite{Faddeev:1996iy}. If
$u_n=iq_n$, then the  Bethe equations in \eqref{ansatz}
take the form
\[\label{betheimag}
\left(\frac{q_n+1/2}{q_n-1/2}\right)^{L}=\prod_{m\neq n}\frac{q_n-q_m+1}{q_n-q_m-1}\,.
\]
For $q_n$ of order one and positive, the exponentially large factor on the 
l.h.s. of \eqref{betheimag}
has to be compensated by a small denominator on the r.h.s., which arises when two
roots are separated by an amount which is close to 1.  Since we are starting
with the roots in \eqref{singbet}, this tends to put the new roots close
to the half-integers.  
We can also see that increasing $L$ 
forces the lower roots closer to the half-integers. 

One can also see that with this configuration of roots $\gamma$ will be of
order
$\order{1/L}$.  If we assume that the roots are equally distributed
along half-integer imaginary values, then from \eqref{anomal} we find
\<
\label{gamapprox}
\gamma\eq\frac{\lambda}{8\pi^2}\left(6-8 \sum_{j=2}^{\frac{J-1}{2}}
\frac{1}{n^2-1}\right)
\nle
\frac{\lambda}{8\pi^2}\left(6-8 \sum_{j=2}^{\infty}
\frac{1}{n^2-1}+8 \sum_{j=\frac{J+1}{2}}^{\infty}
\frac{1}{n^2-1}\right)
\nln
\mathrel{}&\approx&\mathrel{} \frac{8}{\pi^2}\frac{\lambda}{4J}\ .
\>
The prefactor in \eqref{gamapprox} is quite close to the values in
\tabref{tab:halffilled}.
It is also in agreement with the observation of Sec.~\ref{sec:dilop}
that the energy of odd unpaired solutions is 
of order $\order{1/J}$.

\begin{figure}\centering
\includegraphics{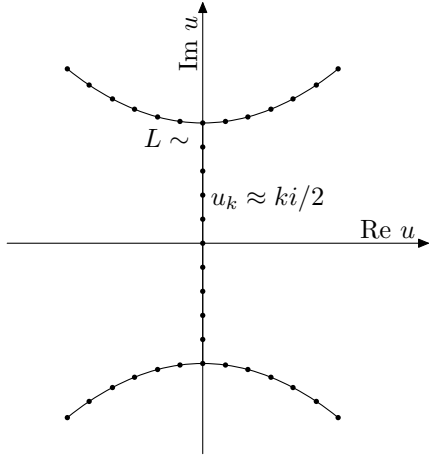}
\caption{Distribution of roots  for the odd ground state.}
\label{fig:oddroots}
\end{figure}

\begin{table}\centering
$\begin{array}{|rc|rc|rc|}\hline
J& \gamma\,[\lambda J/L^2] &J& \gamma\,[\lambda J/L^2] &J& \gamma\,[\lambda J/L^2]\\
\hline
 3&0.911891&13&0.773992&23&0.748305 \\  
 5&0.841644&15&0.766364&  &         \\
 7&0.815506&17&0.760374&  &         \\ 
 9&0.797365&19&0.755558&  &         \\
11&0.783974&21&0.751605&  &         \\\hline
\end{array}$
\caption{Odd ground state energies for $[J,0,J]$}
\label{tab:oddground}
\end{table}

However, for large enough $J$ and $q_n$ we expect this picture 
to break down since the l.h.s. of \eqref{betheimag}  will then be of order 1.   
In fact, for the $[13,0,13]$ case we found that the ground state 
has 9 roots on the imaginary axis and 4
split off from it into the complex plane.  For higher values of $J$,
more roots will break off from the imaginary axis,
see \figref{fig:oddroots}.
In \tabref{tab:oddground} we show the ground state values for $\gamma$ up to
$J=23$.  If we fit the higher data points starting with $J=13$ to the
curve in \eqref{fitfunc} we find
\[\label{betteroddfit}
\gamma=\bigbrk{0.715 + 0.77 /J}\frac{\lambda J}{L^2}\]
indicating that the odd result is asymptoting to \eqref{upshot}.  Hence
the even and odd cases are approaching the same result even though
the configuration of Bethe roots is totally different.

\subsection{States with imaginary roots}\label{sec:imagexp}

Even though they will not give the smallest value for $\gamma$, there still
can be solutions to the Bethe equations with all roots on the imaginary
axis. These remain interesting objects, because they still correspond 
to some string states in \ads.  
What one finds is that the lower roots are very close to the half-integers, 
but the higher roots tend to spread out. In fact, for a given value
of $J>3$, there seem to be multiple solutions of imaginary Bethe roots.
Results for $\gamma$ where the state has all imaginary 
roots are shown in \tabref{tab:FTtable}.  

\begin{table}\centering
$\begin{array}{|c|ccc|}\hline
J&\gamma_{1,J}\,[\lambda J/L^2]&\gamma_{2,J}\,[\lambda J/L^2]&\gamma_{3,J}\,[\lambda J/L^2] \\
\hline
 3&0.911891&        &        \\  
 5&0.841644&1.291309&        \\
 7&0.815506&1.033470&1.684344\\ 
 9&0.797365&0.963623&1.210401\\
11&0.783974&0.931003&1.068883\\
13&        &0.907919&1.008200\\
15&        &0.888306&0.976885\\\hline
\end{array}$
\caption{$[J,0,J]$ states with roots on the imaginary axis}
\label{tab:FTtable}
\end{table}

Let us consider all imaginary solutions
for a given value of $J$, we label them $U_{k,J}$. 
Comparing this to the case $J+2$,
it seems that for each solution $U_{k,J}$ there is a corresponding solution 
$U_{k,J+2}$. In the examples considered in \tabref{tab:FTtable}
this solution has a similar distribution of roots with two additional roots.
Often, these roots are also on the imaginary axis, in some cases, however, they
can spread out into the complex plane. For example, this happens in
$U_{1,13}$, see the discussion in the last subsection.
Although the roots cease to be exclusively on the imaginary axis,
the energy $\gamma_{1,13}$ continues to
follow the extrapolated trajectory, 
see \tabref{tab:oddground}.
However, when we discard these states 
we notice that the value of $\gamma L^2/J$ jumps up,
see \tabref{tab:FTtable}.

We can therefore assume the following picture:
Trajectories of imaginary solutions $U_{k,J}$ appear at some value of $J$.
At a higher value of $J$ some roots of $U_{k,J}$ split off of 
the imaginary axis. 
When we discard such states, which have a relatively low energy, the state with 
the minimum energy will be $U_{k+1,J}$ and the lowest $\gamma L^2/J$ lurches 
upward. 
But one can also see in \tabref{tab:FTtable} that as $J$ becomes large, 
there is some indication that the result is approaching the semiclassical
result of Frolov and Tseytlin for the spinning string.  
We show that this is the case in the following section.

\section{The gauge dual of the semiclassical string}
\label{sec:semiclassical}

In this section, we show that the state with all imaginary roots
has the expected properties of the dual to the semiclassical
string in \cite{Frolov:2003qc}.  We first show that the anomalous dimension
has the expected behavior in the limit of large $L$.  We then consider
spinless fluctuations about the solution and show that they too have
the expected behavior.  

\subsection{A solution for imaginary roots in the large $L$ limit}

In the previous section, we saw evidence that 
if all roots have the form $u_j=iq_j L$ with $q_j$ real, then the roots
close to the real line are very close to the imaginary half integers, but
those further away begin to spread out.  We will call those roots separated
by half integers the ``condensate''.  When $L$ becomes large 
\eqref{logansatz} becomes 
\[
\label{isumeq}
\frac{1}{q_j}=
\frac{2}{L} \sum_{\atopfrac{k=-\frac{J-1}{2}}{k \neq j}}^{\frac{J-1}{2}}
\frac{1}{q_j - q_k}\ 
\]
where \eqref{isumeq} is valid for $q_j$ outside the condensed region.
Note in particular that there is no integer on the r.h.s. due to 
a branch ambiguity.  The equation in \eqref{isumeq} can be rewritten
as an integral equation
\[
\label{iinteq}
\frac{1}{q}=\pint dq'\,\frac{\s(q')}{q-q'}\ ,
\]
which is valid for $q$ outside the condensate and 
where the root density $\s(q')$ is given by 
\[
\label{rootd}
\s(q')=\frac{2}{L}\sum_ {k=-\frac{J-1}{2}}^{\frac{J-1}{2}}\delta(q'-q_k)\ .
\]

We now make the ansatz that the roots are condensed in
the interior region, but start spreading out at some point of order $L$
from the origin.  This means that the root density is of the form
\begin{equation}
\label{rootden}
\s(q)\ =\ \left\{   
\begin{array}{ll}
4 &\qquad -s<q<s, \\
  \ts(q) &\qquad  s<q<t, \\
\ts(-q)  &\qquad -t<q<-s, \\
    0 &\qquad q<-t\ \ {\rm or}\ \ t<q
\end{array}
\right.
\end{equation}
where $t$ and $s$ are to be determined.  Notice that this configuration
is very similar to one considered by Douglas and Kazakov for the strongly
coupled phase of ${\rm QCD}_2$ on the sphere \cite{Douglas:1993ii}, 
where the density
also has a condensed region.  The problem considered
here is somewhat different since the ${\rm QCD}_2$ equations would have
$q$ on the l.h.s. of
\eqref{iinteq} instead of $1/q$. In terms of $\ts(q)$, \eqref{iinteq}
becomes
\[
\label{newinteq}
\frac{1}{q}-4\ln\frac{q+s}{q-s}
=\pint_s^t  dq'\,\ts(q')\left(\frac{1}{q-q'}+\frac{1}{q+q'}\right)\ .
\]
The solution of this integral equation can be written in the form
\[
\label{tseq}
\ts(q)=\frac{1}{\pi}\sqrt{(q^2-s^2)(t^2-q^2)}\left(-\frac{1}{qst}
+4\int_{-s}^s\frac{dv}{(q-v)\sqrt{(s^2-v^2)(t^2-v^2)}}\right)\ .
\]

If we plug \eqref{tseq} back into \eqref{newinteq}, we see that $s$ and
$t$ must satisfy a consistency condition
\[
\label{conscon}
4\int_{-s}^s\frac{dv}{\sqrt{(s^2-v^2)(t^2-v^2)}}=\frac{1}{st}\ .
\]
We also have the normalization condition for $\s(q)$
\[
\label{normcond}
\int dq\ \s(q)=8s+2\int_s^t dq\ \ts(q)=2\al\ ,
\]
where $\al$ is the previously defined filling factor.  If we now insert
\eqref{tseq} into \eqref{normcond} and use 
\eqref{conscon} we find the equation
\[
\label{critcond}
4\int_{-s}^s\frac{dv\ v^2}{\sqrt{(s^2-v^2)(t^2-v^2)}}=1-2\alpha\ .
\]

The anomalous dimension is found by using \eqref{anomal}, \eqref{gamapprox}
 and \eqref{rootden}, which gives
\<
\label{gamieq}
\gamma\eq\frac{\la}{8\pi^2}\left(6-2\sum_{n=2}^{2sL}\frac{4}{n^2-1}-
2\sum_{n=sL}^\infty\frac{1}{q_n^2}\right)+\order{1/L^2}
\nle
\frac{\la}{8\pi^2}\left(\frac{4}{sL}-\frac{1}{L}\int_s^tdv\frac{\ts(v)}{v^2}
\right)+\order{1/L^2}\ .
\>
The integral inside \eqref{gamieq} is found by 
using \eqref{tseq}, \eqref{conscon} and \eqref{critcond}, which leads
to
\[
\label{gammast}
\gamma=\frac{\la}{32\pi^2L}\left(\frac{1}{s^2}+\frac{1}{t^2}-2\ \frac{1-2\al}{st}\right)\ .
\]

Let us now focus on the limiting case $\al=1/2$ which corresponds to the
$[J,0,J]$ representation.  Using
\eqref{critcond} and \eqref{conscon}, we see that in the limit 
$\al\to 1/2$, $t$ and $s$ approach
\[
t\to\infty\qquad\qquad s\to\frac{1}{4\pi}\ .
\]
Therefore, setting $\al=1/2$, we find
\[
\label{gammaFT}
\gamma=\frac{\lambda}{2L}=\frac{\lambda}{4J}\ ,
\]
which is precisely the string theory prediction in \cite{Frolov:2003qc}.
Starting from this result we can also expand the energy around $\alpha=1/2$. 
We get 
\[
\gamma=
\frac{\lambda}{2L}\bigbrk{1
+8(\half-\alpha)^2
+24(\half-\alpha)^4
+96(\half-\alpha)^6
+408(\half-\alpha)^8
+\ldots
}
\]
Interestingly, the energy is symmetric around
$\alpha=1/2$ and reaches a maximum at $\alpha=1/2$.

We can also see that the solutions for the $[J,L-2J,J]$ representations
do not approach a BMN limit when $J\ll L$.  The easiest way to see
this is to realize that the roots are all close to imaginary half-integers and
so we can borrow the result in \eqref{gamapprox}, 
giving\footnote{The same results can  be derived from eqs.~\eqref{conscon},
\eqref{normcond}, \eqref{gammast} by noticing that $s\approx t\approx \alpha/4$
and $t-s\sim\exp(-4/\alpha^2)$ as $\alpha\rightarrow 0$.} 
$\gamma=2\la/\pi^2J$
if $J\ll L$.  Hence, for finite $J$ the anomalous dimension is finite in
the large $L$ limit.

Note also that for
$\al=1/2$, the root density simplifies to
\[
\label{simprd}
\ts(q)=4\left(1-\sqrt{1-\frac{1}{(4\pi q)^2}}\right)\ .
\]
Hence, $4-\ts(q)$ has a Wigner distribution in $1/q$.  This is like
the critical point for Douglas-Kazakov, where the eigenvalue distribution
changes to a  Wigner distribution at the critical area.

\subsection{Fluctuations}

We can also show that the fluctuations of the solution in the previous
section are consistent with the Frolov-Tseytlin prediction.
The fluctuations we consider are spinless,
that is, they do not change the $\grSO(6)$ representation.  Hence, these
fluctuations are found by moving Bethe roots around in the complex plane, 
but not actually changing the net total of roots.

With this in mind, let us 
suppose that two roots are moved from the imaginary axis.
If we assume that the remaining roots stay
on the axis then in order to satisfy the trace condition, the two roots
must be at $\pm\mu$ where $\mu$ is positive real.  
In the large $L$ limit, $\mu$
satisfies the equation
\[
\frac{1}{\mu}=2\pi n + \int dq\,\frac{\s(q)}{\mu-iq}\ ,
\]
where we have again rescaled the roots by a factor of $L$.  The integer $n$
corresponds to the different branch choice for the log.  
The root at $-\mu$ is on the $-n$ branch.   We may also
assume that the function $\s(q)$ is the same one computed in the previous
section.  Using $\s(q)=\s(-q)$ and the expression in \eqref{simprd}, we
obtain the equation
\[
\label{fluceq}
\frac{1}{\mu}=2\pi n+\int_{-\infty}^{\infty}dq\,\frac{4\mu}{\mu^2+q^2}
-4\mu\left(\int_{-\infty}^{-s}dq+\int_{s}^{\infty}dq\right)
\frac{\sqrt{q^2-s^2}}{q(q^2+\mu^2)}
\]
Deforming the contour, we then find
\[
\label{musol}
\mu^{-1}=2\pi\sqrt{n(n+4)}\ .
\]
Hence, we see that $n>0$  for $\mu$ to be positive real.
The contribution of these roots to the anomalous dimension is
\[
\label{gammaroots}
\gamma\indup{\mu}=\frac{2\lambda}{8\pi^2 L^2\mu^2}=\frac{n(n+4)\lambda}{L^2}\ .
\]

The roots at $\pm\mu$ also back react on the imaginary axis roots.
To find this effect, we note that the roots at $\mu$ modify the equation in
\eqref{newinteq} to
\[
\label{eqwimp}
\frac{1}{q}-4\ln\frac{q+s}{q-s}-\frac{4q}{L(q^2+\mu^2)}
=\pint_s^t dq'\,\ts(q')\left(\frac{1}{q-q'}+\frac{1}{q+q'}\right)\ .
\]
We can then solve for $\ts(q)$ with the new term on the l.h.s. of the equation
in \eqref{eqwimp}.  Thus, we have
\[
\label{normcond2}
\int dq\ \s(q)=8s+2\int_s^t dq\ \ts(q)=2\al\ -\frac{4}{L},
\]
\[
\label{conscon2}
4\int_{-s}^s\frac{dv}{\sqrt{(s^2-v^2)(t^2-v^2)}}=\frac{1}{st}\ -
\frac{4}{L}\frac{1}{\sqrt{(s^2+\mu^2)(t^2+\mu^2)}}\ ,
\]
and the same equation in \eqref{critcond}.
Hence, we see that the distribution still has $t\to\infty$ if $\al=1/2$, but
the position of s is shifted to
\[
\label{sapprox}
s\approx\frac{1}{4\pi}\left(1-\frac{4}{L}\frac{1}{\sqrt{1+(4\pi\mu)^2}}\right).
\]  

We now compute the contribution of the imaginary roots to the anomalous dimension,  where we find
\[
\label{gammacond}
\gamma\indup{ir}=\frac{\la}{32\pi^2L}\left(\frac{1}{s^2}+\frac{1}{t^2}-2\ \frac{1-2\al}{st}-\frac{2}{L\mu}\left(1-\frac{st}{\sqrt{(s^2+\mu^2)(t^2+\mu^2)}}\right)\right)\ .
\]
If we now let $\al=1/2$, use  \eqref{musol} and \eqref{sapprox}, and then
add \eqref{gammaroots} to \eqref{gammacond}, we find that the change
in the anomalous dimension from \eqref{gammaFT}
is 
\[
\label{Deltagamma}
\Delta\gamma=\frac{\lambda}{L^2}(n+2)\sqrt{n(n+4)}\ .
\]
This is the prediction of Frolov and Tseytlin, where our definition of
$n$ is shifted by 2 from theirs,
and with extra factor of 2 since this configuration is essentially two
fluctuation quanta.

In their semiclassical analysis, Frolov and Tseytlin also identified an
unstable mode, that is, a mode with an imaginary frequency.  
Since we are considering eigenstates of a Hermitian
Hamiltonian we will not see this directly.  It would seem that
$\Delta\gamma$ is imaginary if $n=-1$ or $-3$.  However, in this
case $\mu$ would also be imaginary.  But the derivation of \eqref{musol}
assumed that $\Re(\mu)>0$.  Hence the imaginary $\mu$ solution
is not actually a solution to the Bethe equations.

On the other hand, we have already established that there are states
with lower energy and in the large $L$ limit we would expect them to form
a continuum, see Sec.~\ref{sec:imagexp}.
Hence there can be a superposition of states centered about
the imaginary root solution that will decay.   If the superposition is
spread out over solutions with width $\order{1/L^{2}}$, then this will
lead to frequencies with imaginary parts of the same order.

Even though the imaginary $\Delta\gamma$ do not correspond to solutions to the
Bethe equations, they still give the predicted imaginary frequencies of
Frolov and Tseytlin.  Hence a possible interpretation is that the states
constructed with these values of $\mu$ are precisely these superpositions
of energy eigenstates.

Placing roots at $\pm\mu$ is consistent with level matching of
the fluctuations.  
If we were to consider a finite number of
roots removed from the axis and placed at $\pm\mu_i$, 
we may ignore their interactions among
themselves and so the
change in the anomalous dimension is additive.  If we were to put two
roots on the same cut $n$, they would split off from the real line
\cite{Minahan:2002ve}, but
the splitting is of order $1/\sqrt{L}$ and so this can also be ignored
at order $1/L^2$.   

{}From the string perspective, there can be
nonsymmetric configurations of roots that are still consistent with
level matching.   Assuming that such states exist,
 then the imaginary roots will also be nonsymmetric, 
which means that they must
stray from the imaginary axis.  
It would be interesting to find these solutions.
Also it would be interesting to find 
an analog of the semiclassical string configuration for even $J$.

\section{An $\grSO(6)$ singlet solution}
\label{sec:singlet}

It turns out that we can find a solution for an $\grSO(6)$ singlet that
is very similar to the solution in the last section.  Once we
consider the full $\grSO(6)$, then it is necessary to introduce two
other types of Bethe roots \cite{Minahan:2002ve}.    If we assume
that all roots are imaginary and that the distribution of the two new
sets of roots are the same, then the full $\grSO(6)$ Bethe equations
become
\<
\left(\frac{q_n+1/2}{q_n-1/2}\right)^L\eq\prod_{m\ne n}\frac{q_n-q_m+1}{q_n-q_m-1}\prod_k\left(\frac{q_n-r_k-1/2}{q_n-r_k+1/2}\right)^2
\nln
1\eq\prod_{k\ne j}\frac{r_j-r_k+1}{r_j-r_k-1}\prod_m\frac{r_j-q_m-1/2}{r_j-q_m+1/2}
\>
where $r_j$ refer to the new set of roots.  Taking logs, rescaling by $L$
 and converting
these equations to integral equations, we find
\<
\label{so6beq}
\frac{1}{q}\eq\pint dv\,\frac{\s(v)}{q-v}-\pint dv\, \frac{\omega(v)}{q-v}
\nln
0\eq\pint dv\, \frac{\omega(v)}{r-v}-\frac{1}{2}\pint dv\,\frac{\s(v)}{r-v}
\>
where $\s(q)$ is the same as \eqref{rootd}
and $\omega(q)$ is
\[
\omega(r)=\frac{2}{L}\sum_k\delta(r-r_k)\ .
\]
Therefore, we have
\[
\frac{2}{q}=\pint dv\,\frac{\s(v)}{q-v}
\]
and so we have almost the same equation as in the previous section.  
The factor of $2$ leads to the modified equations
\<
\label{newcritc}
4\int_{-s}^s\frac{dv}{\sqrt{(s^2-v^2)(t^2-v^2)}}\ =\ \frac{2}{st}\ 
\nln
4\int_{-s}^s\frac{dv\ v^2}{\sqrt{(s^2-v^2)(t^2-v^2)}}\ =\ 2-2\alpha\ 
\nln
\gamma=\frac{\la}{16\pi^2L}\left(\frac{1}{s^2}+\frac{1}{t^2}-2\ \frac{1-\al}{st}\right)\ .
\>

In the $\grSO(6)$ integrable chain
there can be as many as $L$ of the $q$ type roots
and $L/2$ of the $r$ type.  In this case, the representation is a singlet.
Hence we see that the maximum filling fraction
is $\al=1$, which is reflected in the r.h.s. of \eqref{newcritc}.  We
also see that $s=1/(2\pi)$ when $\al=1$. Therefore, the anomalous dimension
is
\[
\label{so6ad}
\gamma=\frac{1}{4L}\ ,
\]
which matches the semiclassical prediction for a circular string
pulsating on $S_5$ \cite{Minahan:2002rc}.

\section{Conclusions}
\label{sec:concl}

In this paper we have constructed two types of gauge invariant operators
which are in $[J,L-2J,J]$ representations.  The first type is 
a limit of BMN operators and was shown to have the smallest
anomalous dimension for this representation.  The second type of solution
had a higher anomalous dimension, but was shown to have the same anomalous
dimension and fluctuation spectrum as the semiclassical string, strongly
suggesting that this is the gauge dual.

The fact that the 
semiclassical string result in \cite{Frolov:2003qc} does not correspond
to the operator with smallest anomalous dimension is mildly surprising;
one usually expects a state with a large amount of symmetry to have
a low, if not the lowest energy.
It suggests that there are other semiclassical solutions to be found.
It also suggests that these semiclassical 
solutions will involve elliptic integrals
in order to reproduce the anomalous dimension.

It would be very important, in order to check and complete our ideas,
to find the solution of the Bethe integral equation for the ground state in
the case of odd $J$, \emph{c.f.}~section~\ref{sec:oddground}.  
Likewise, it would be
exciting to find the analog of the purely imaginary solutions
for even $J$. On physical grounds, we expect no difference
in the energies between even and odd for large $J$, but as we have seen,
the root distributions are certainly very different.

The computations presented here
are one-loop calculations, but nevertheless not ones
that would be accessible to standard Feynman diagram and combinatorial
techniques, as the real difficulty is the diagonalization of the 
states.
This nicely illustrates the power of using the Bethe ansatz to
do one-loop computations of anomalous
dimensions in SYM and suggests that it might be applicable to
even more gauge invariant operators. 
It also illustrates the extremely rich mathematical
structure hidden in $\superN =4$.

Finally, let us
stress the fact that the problems solved in this paper are
quite different from the ones considered by condensed matter theorists.
For magnetic chains one is either
interested in the ferromagnetic or the antiferromagnetic phase.
The Bethe vacuum (corresponding to the BPS states in 
gauge theory) is the true vacuum of the ferromagnet, 
but only a ``fake" vacuum of the antiferromagnet.
For finite filling fraction $\alpha$, the two-contour states we considered are
hybrids -- they are ``antiferromagnetic" in the sense that 
they have low spin,
but ferromagnetic in the sense
that they have the smallest 
energy, {\it i.e.}  anomalous dimension, for fixed $\alpha$. 
For this reason and 
to the best of our knowledge, despite all the work on the XXX Heisenberg chain 
since its invention more than 70 years ago,
this problem has
not previously been solved.

{\bf Note Added:}  After this paper was completed we learned that Frolov
and Tseytlin have studied fluctuations around the stable 
counterparts of the string soliton dual to the Bethe states 
discussed here.  They found that the $\order{\lambda/L}$ term
in the anomalous dimension is indeed
not renormalized by the string 
corrections \cite{Frolov:2003xx}. 

{\bf Second Note Added:} After this paper was submitted to the arXiv
we were informed that Frolov and Tseytlin  were able to
find classical string configurations
which are dual to the double contour solution presented here.
\cite{FTnew}.

\subsection*{Acknowledgments}

We are grateful to S.~Frolov and A.~Tseytlin for correspondence.
N.B.~and M.S.~would like to thank the Department
of Theoretical Physics in Uppsala for hospitality. 
J.A.M. would like to thank the CTP at MIT for hospitality.
The work of J.A.M. and K.Z. was
supported in part by the Swedish Research Council.
The work of K.Z. was also supported in part by
RFBR grant  01-01-00549 and in part by  grant
00-15-96557 for the promotion of scientific schools.
N.B.~dankt der \emph{Studienstiftung des
deutschen Volkes} f\"ur die Unterst\"utzung durch ein 
Promotions\-f\"orderungsstipendium.


\newpage
\section*{Tables}

\begin{table}[h]\centering
$\begin{array}{|c|lll|c|}\hline
\gamma \,[\lambda J/L^2]& \multicolumn{3}{c|}{\mbox{Bethe roots}}                & n \\\hline
0.675927& \multicolumn{2}{c}{(\pm 0.676245\pm 0.993633i)}& \pm 0.678017  & 1\,1\,1\\
0.948774& \multicolumn{2}{c}{(\pm 0.120258\pm 0.500000i)}& \pm 1.638344i & 2\,2\,0\\
1.485952& \pm 2.177728i& \pm 1.000078i& \pm 0.001879                     & 0\,0\,5\\
1.913430& \multicolumn{2}{c}{(\pm 0.347918\pm 0.500019i)}& \pm 0.695172  & 2\,2\,1\\
2.303250& \pm 1.003317i& \pm 0.805340 & \pm 0.012646                     & 0\,1\,5\\
2.804933& \multicolumn{2}{c}{(\pm 0.519954\pm 0.500990i)}& \pm 0.279477  & 1\,1\,4\\
3.163442& \pm 1.012980i& \pm 0.386936 & \pm 0.024593                     & 0\,2\,5\\
3.283924& \multicolumn{2}{c}{(\pm 0.560191\pm 0.501721i)}& \pm 0.083335  & 1\,1\,5\\
3.854384& \pm 1.063338i& \pm 0.186496 & \pm 0.047508                     & 0\,3\,5\\
5.098922& \pm 0.657299 & \pm 0.282367 & \pm 0.084694                     & 1\,3\,5\\
\hline
\end{array}$
\caption{Unpaired states of $[6,0,6]$}
\label{tab:yetanothertable606}
\end{table}

\begin{table}[h]\centering
$\begin{array}{|c|llll|c|}\hline
\gamma \,[\lambda J/L^2] & \multicolumn{4}{c|}{\mbox{Bethe roots}}             & n \\\hline
0.815506& 0&\pm i/2&\pm 1.000016i &\pm 1.478037i & 6\,\mathord{\ast}\,1\,1\\
1.033467& 0&\pm i/2&\pm 0.999996i &\pm 2.241154i & 5\,\mathord{\ast}\,1\,0\\
1.684344& 0&\pm i/2&\pm 1.500899i &\pm 2.862938i & 6\,\mathord{\ast}\,0\,0\\
1.870623& 0&\pm i/2&\pm 0.999885i &\pm 0.882971  & 5\,\mathord{\ast}\,1\,1\\ 
2.366246& 0&\pm i/2&\pm 1.524153i &\pm 0.985655  & 6\,\mathord{\ast}\,0\,1\\ 
2.749876& 0&\pm i/2&\pm 0.999612i &\pm 0.449560  & 5\,\mathord{\ast}\,1\,2\\ 
3.065353& 0&\pm i/2&\multicolumn{2}{c|}{(\pm 0.701060 \pm 0.504190i)}  & 6\,\mathord{\ast}\,1\,1\\ 
3.156982& 0&\pm i/2&\pm 1.555226i &\pm 0.522436  & 6\,\mathord{\ast}\,0\,2\\ 
3.514967& 0&\pm i/2&\pm 0.998615i &\pm 0.231443  & 5\,\mathord{\ast}\,1\,3\\ 
3.915465& 0&\pm i/2&\pm 1.584609i &\pm 0.295822  & 6\,\mathord{\ast}\,0\,3\\ 
3.939509& 0&\pm i/2&\pm 0.986006i &\pm 0.062156  & 5\,\mathord{\ast}\,1\,4\\ 
4.462322& 0&\pm i/2&\pm 1.603217i &\pm 0.136490  & 6\,\mathord{\ast}\,0\,4\\ 
4.926542& 0&\pm i/2&\pm 0.740432  &\pm 0.348252  & 6\,\mathord{\ast}\,1\,3\\ 
5.587007& 0&\pm i/2&\pm 0.752202  &\pm 0.154446  & 6\,\mathord{\ast}\,1\,4\\ 
6.559173& 0&\pm i/2&\pm 0.364262  &\pm 0.158450  & 6\,\mathord{\ast}\,2\,4\\ 
\hline
\end{array}$
\caption{Unpaired states of $[7,0,7]$}
\label{tab:yetanothertable707}
\end{table}

\end{document}